\documentstyle[12pt]{article}
\newcommand\be{\begin{equation}}
\newcommand\ee{\end{equation}}
\newcommand\bea{\begin{eqnarray}}
\newcommand\eea{\end{eqnarray}}

\newcommand{\fatalpha}{{\bf \alpha \kern -0.44em \alpha}}
\newcommand{\fatsigma}{{\bf \sigma \kern -0.54em \sigma}}
\newcommand{\tpchi}{{\bf \chi \kern -0.35em \chi}}
\newcommand{\llambda}{{\bf \lambda \kern -0.45em \lambda}}

\renewcommand{\theequation}{\arabic{equation}}
\renewcommand{\theequation}{\thesection.\arabic{equation}}
\bibliography{plain}
\title{\bf  Generic N-coupled maps in
Bose-Mesner algebra perspective } 
\author{ M. A. Jafarizadeh$^{a,b,c}$
 \thanks{E-mail:jafarizadeh@tabrizu.ac.ir}, S.Behnia$^{d}$\thanks{E-mail:s.behnia@iaurmia.ac.ir},
 E.Faizi$^{a,c}$\thanks{E-mail:efaizi@tabrizu.ac.ir} and S. Ahadpour$^{a,c}$\thanks{E-mail:ahadpour@tabrizu.ac.ir}\\
$^a${\small Department of Theoretical Physics and Astrophysics,
Tabriz University, Tabriz 51664, Iran.} \\ $^b${\small Institute
for Studies in Theoretical Physics and Mathematics, Tehran
19395-1795, Iran.} \\ $^c${\small Research Institute for
Fundamental Sciences, Tabriz 51664, Iran.}\\$^d ${\small
Department of Physics, Islamic Azad University, Urmia, Iran.}}
\pagebreak
\begin{document}
\maketitle
\newpage 
\begin{abstract}
By choosing a dynamical system with {\bf d} different couplings,
one can rearrange a system based on the graph with given vertex
dependent on the dynamical system elements.  The relation between
the dynamical elements (coupling) is replaced by a relation
between the vertexes. Based on the $E_0$ transverse projection
operator. We addressed synchronization problem of an array of the
linearly coupled map lattices of identical discrete time systems.
 The
synchronization rate is determined by the second largest
eigenvalue of the transition probability matrix. Algebraic
properties of the Bose-Mesner algebra with an associated scheme
with definite spectrum has been used in order to study the
stability of the coupled map lattice. Associated schemes play a
key role and may lead to analytical methods in studying the
stability of the dynamical systems. The relation between the
coupling parameters and the chaotic region is presented.  It is
shown that the feasible region is analytically determined by the
number of couplings (i.e, by increasing the number of coupled maps
the feasible region is restricted). It is very easy to apply our
criteria to the system being studied and they encompass a wide
range of coupling schemes including most of the popularly used
ones in the literature.
\end{abstract}

{\bf PACS index. 05.45.Xt,84.35.Xt}
\newpage
\section{Introduction}
Simulation of the natural phenomena is one of the most important
research fields and coupled map lattices are a paradigm for
studying fundamental questions in spatially extended dynamical
systems. The early definition of the coupled map lattice goes back
to the Kaneko's paper\cite{In-1}.  Now we can divide them in two
categories: internal and external coupled map lattice\cite{In-2},
Globally coupled map is one of the most well know examples of the
external coupling, with different number of elements\cite{In-3}.
Variety of  couplings  such as the week coupling\cite{In-4}, noisy
coupling\cite{In-5} and functional coupling\cite{In-6}
used to made the new coupled map.\\
On the other hand many areas of research lie on the interface of
algebra and physics. Bose-Mesner algebra arose independently in
three areas: statistical designs, centralizer algebras of
permutation groups, distance-transitive graphs  and quantum walk
\cite{S1-0, S1-1, S1-2, S1-21, S1-22}. Associated schemes are
combinatorial objects that allow us solve problems in several
branches of mathematics. They have been used in the study of
permutation groups and graphs and also in the design of
experiments, coding theory, partition designs etc.
 In this paper the authors are interested to use the ability of
algebraic properties of the Bose-Mesner algebra associated with an
association scheme with known spectrum in order to study the
stability of the coupled map lattice, since elements of
Bose-Mesner algebra have capability that can
be diagonalized simultaneously. \\
 The aim of the present paper is
threefold: introducing a generalized model of the coupled map
lattice which covers internal and external couplings in a form of
associated schemes. By taking into account that idempotents are
another basis of this algebra ($E_o$ longitude projection
operator, $E_{\beta}$ transverse projection operator),
 in our proposed model all of the elements have a common
prolongation.  Idempotents are independent of coupling and
coupling method influences the
stability of the coupled map lattices.\\
We addressed synchronization problem of an array of the linearly
coupled map lattices of identical discrete-time systems. We
present the proof that the synchronization rate is mainly
determined by the second largest eigenvalue of the Floquet
multipliers.
 We also prove that topological behavior can be
studied by Sinai-Bowen-Rule (SBR) measure for the coupled map
lattice of measurable dynamical system at the complete
synchronized state.\\
 In order to detect the feasible region, the stability analysis of
the coupled map lattice  with regard to the Lyapunov exponent is
possible. We define the Lyapunov function by refereing to the
projection operators $E_{\alpha}$'s in order to confirm the
stability analysis based on the Lyapunov exponents. Lyapunov
function takes 0 value at synchronized state. We have investigated
some conditions on transition from regularity to chaos with
Li-yorke theorem in a coupled map network, whose individual nodes
are non-chaotic before being connected to the network. It has been
discovered that, such a transition to chaos depends not only on
the network topology but
also on the original node dynamics.\\
In order to determine whether our results in this paper are
generic or they are only limited to the specific model, we study
several other coupled maps. It was proved that the global coupled
map lattice corresponds to the compete graph; and the nearest
neighbor coupled map can be studied based on the model of
Associated schemes.  It is very easy to apply our criteria to the
system being studied, and they encompass a wide range of coupling
schemes including most of the popularly used
ones in the literature.\\
 The paper is organized as follows: In section 2 of the paper we
present a brief outline of some of the main features of the
associated schemes, such as adjacency matrix, distance-regular
graphs, stratification and orthonormal basis of the strata. In
Section 3, the generic model of the discrete-time coupled map
Lattice (Internal and coupled map lattice) is introduced in
adjacency matric perspective. In section 3, the synchronization
conditions are also studied. Section 4, discusses the existence of
SBR-measure at the synchronized state in the light of a measurable
dynamical system. In the section 5, the stability analysis of the
coupled map lattice based on the discussion of Floquet multipliers
of coupled map lattice is done through three subsections; Lyapunov
exponent calculation, Lyaunov function analysis and Li-yorke
theorem. In section 6, we present some examples of the coupled map
lattice model and their corespondent graph such as the complete
graph, strongly regular graphs and cycle graph. This section
followed by an outlook section. Respective appendix can be found
at the end of paper.
\section{Association schemes}
In this section we give a brief outline of some of the main
features of association scheme, and the reader is referred to
\cite{S1-0}for further information on the Associated schemes.
Recall that a finite graph is a  finite set $\Gamma$, whose
elements are called vertices described by vertex set $V$, together
with a set of $2-subset$ of $\Gamma$ called edges.  An association
scheme with d associate classes on a finite set $\Gamma$ is
coloring of all the edges of the graph  by s colors such that the
set of edges with the same color form a non empty relation subsets
on $\Gamma$ (i,e, subset of $ \Gamma\times \Gamma$ ), denoted by
$R_i$, where $i=1,..,d$ corresponds to different colors. Then the
pair $Y=(\Gamma, \{R_i\}_{0\leq i\leq d})$ consisting of a set
$\Gamma$ and a set of relations $\{R_i\}_{0\leq i\leq d}$ together
with the following 4 conditions is called an association scheme.
\begin{description}
  \item [1-]$ \{R_i\}_{0\leq i\leq d}$ is a partition of $\Gamma\times
\Gamma$
  \item [2-] $ R_0=\{(\alpha, \alpha) : \alpha\in \Gamma \}$
  \item [3-]$ R_i=R_i^t$ for $0\leq i\leq d$, where
$R_i^t=\{(\beta,\alpha) : (\alpha, \beta)\in R_i\} $
\item [4-] Given $(\alpha, \beta)\in R_k, p_{ij}^{k}=\mid \{\gamma\in
\Gamma : (\alpha, \gamma)\in R_i \;\ and \;\ (\gamma,\beta)\in
R_j\}\mid$,
\end{description}
where the constants $p_{ij}^{k}$ are called the intersection
numbers, depending only on $i, j$ and $k$ and not on the choice of
$(\alpha, \beta)\in R_k$. Then the number N of the vertices V is
called the order of the association scheme and $R_i$ is called a
relation or associate class. Let $\Gamma=(V,R)$ denote a finite,
undirected, connected graph, with vertex set V edge set R
path-length distance function $\partial$, and diameter
$d:=max\{\partial(\alpha,\beta):\alpha,\beta \in V\}$. For all
$\alpha$,$\beta$ $ \in V$ and all integer i ,we set $
\Gamma_{i}=(V,R_{i})=\{(\alpha,\beta):\alpha,\beta \in V
:\partial(\alpha,\beta)=i \}$ so that $\Gamma_{i}(\alpha)=\{\beta
\in V :\partial(\alpha,\beta)=i \}$.

\section{The Bose-Mesner algebra} Let $R$
denote the field of complex numbers. By $Mat_{V}(R)$ we mean the
$R$-algebra consisting of all matrices whose entries are in $R$
and whose rows and columns are indexed by $V$. For each integer
$i$ ($0 \leq \alpha\leq d$), let $A_\alpha$ denote the matrix in
$Mat_V (R)$ with $(i, j)$-entry
\begin{equation}
\bigl(A_{\alpha})_{i, j}\;=\;\cases{1 & if $\;(i, j)\in
R_\alpha$,\cr 0 & otherwise\cr}\qquad \qquad (i, j \in \Gamma).
\end{equation}
The matrix $A_\alpha$ is called an adjacency matrix of the
association scheme \cite{S1-1}. Then we have $ A_0=I$(by (2)
above) and
\begin{equation}\label{li}
A_\alpha A_\beta=\sum_{\gamma=0}^{d}p_{\alpha\beta}^\gamma
A_{\gamma},
\end{equation}
(by (4) above), so $A_0, A_1, ..., A_d$ form a basis for a
commutative algebra \textsf{A} of $Mat_V(C)$, where \textsf{A} is
known as the Bose-Mesner algebra of $Y$. Since the matrices $A_i$
commute, they can be diagonalized simultaneously, that is, there
exists a matrix $S$ such that for each $A\in \textsf{A}$,
$S^{-1}AS$ is a diagonal matrix. Therefore $\textsf{A}$ is
semi-simple and has a second basis $E_0,..., E_d$ (see
\cite{S1-2}). These are matrices satisfying
\begin{equation}
E_0 = \frac{1}{N}J_{N},\quad E_\alpha
E_\beta=\delta_{\alpha\beta}E_\alpha,\quad \sum_{\alpha=0}^d
E_\alpha=I_{N}.
\end{equation}
The matrix $\frac{1}{N}J_{N}$ (where $J_{N}$ is the all-one matrix
in $\textsf{A}$ and $N=\mid\Gamma\mid$ ) is a minimal idempotent
(idempotent is clear, and minimal follows since rank matrix
$J_{N}$ equal  1. The $E_\alpha$, for ($0\leq \alpha,\beta\leq d$)
are known as the primitive idempotent of $Y$. Let $P$ and $Q$ be
the matrices relating our two bases for $\textsf{A}$:
\begin{equation}
A_\beta=\sum_{\alpha=0}^d P_{\alpha\beta}E_\alpha, \;\;\;\;\ 0\leq
\beta\leq d,
\end{equation}
\begin{equation}\label{m2}
E_\beta=\frac{1}{N}\sum_{\alpha=0}^d Q_{\alpha\beta}A_\alpha,
\;\;\;\;\ 0\leq \beta\leq d.
\end{equation}
Then clearly
\begin{equation}
PQ=QP=NI_{N}.
\end{equation}
It also follows that
\begin{equation}
A_\beta E_\alpha=P_{\alpha\beta}E_\alpha,
\end{equation}
which shows that the $P_{\alpha\beta}$ (resp. $Q_{\alpha\beta}$)
is the $\alpha$-th eigenvalues(resp. the $\alpha$-th dual
eigenvalues ) of $A_\beta$ (resp. $E_\beta$) and that the columns
of $E_\alpha$ are the corresponding eigenvectors. Thus $m_\alpha=$
rank($E_\alpha$) is the multiplicity of the eigenvalue
$P_{\alpha\beta}$ of $A_\beta$ (provided that $P_{\alpha\beta}\neq
P_{\gamma\beta}$ for $\gamma \neq \alpha$). We see that $m_0=1,
\sum_\alpha m_\alpha=N$, and
$m_\alpha=$trace$E_\alpha=N(E_\alpha)_{\beta\beta}$ (indeed,
$E_\alpha$ has only eigenvalues $0$ and $1$, so rank($E_\gamma$)
equals the sum of the eigenvalues). Also,
  the eigenvalues and dual eigenvalues satisfy
 \begin{equation}
 m_\beta P_{\beta \alpha}=k_\alpha Q_{\alpha \beta}, \;\;\;\;\ 0\leq
\alpha,\beta\leq d,
 \end{equation}
 where for all integer $\alpha$ ($0\leq \alpha\leq d$), set
 $k_\alpha=p_{\alpha\alpha}^{0}=(N-1)^{\alpha}C_{N}^{\alpha}$, and note
that $k_\alpha\neq 0$, since $R_\alpha$ is
 non-empty.
\subsection{Definition of generic N-coupled map}
Coupled map lattices are arrays of states whose values are
continuous, usually within the unit interval, or discrete space
and time. The generic N-coupled map dynamical system can be
considered as a N-dimensional dynamical  map defined as:
\begin{equation}
\Phi(x_{1}(n),...,x_{N}(n))=\left\{\begin{array}{l}
x_{1}(n+1)=F_1(x_{1}(n),...,x_{N}(n))\\x_{2}(n+1)=F_2(x_{1}(n),...,x_{N}(n))\\\vdots\\x_{N}(n+1)=F_N(x_{1}(n),...,x_{N}(n)).
\end{array}\right.
\end{equation}
Adjacency matrix element such as $A_0$  can cover the elements of
the dynamical systems in the coupled map lattice.  $A_1$, $A_2$,
etc, have the capability of generating the different coupled map
matrices such as nearest neighbor coupled map, second nearest
neighbor coupled map and globally coupled maps.
  Definition of the coupled
map model based on the adjacency matrix allows us to use the
spectrum of graph based on the associated scheme.
 Therefore, generic N-coupled map is defined as
follows:
\begin{equation}
x_i(n+1)=\left[\epsilon
\sum_{\alpha=0}^{d}\frac{\epsilon_{\alpha}^{'}}{k_{\alpha}}\left(\sum_{x\in\Gamma_{\alpha}(x_{i}(n))}\Phi(x(n)
)\right)+(1-\epsilon)\Phi\left(\sum_{\alpha=0}^{d}\frac{\epsilon_{\alpha}^{''}}
{k_{\alpha}}(\sum_{x\in\Gamma_{\alpha}(x_{i}(n))}x(n))\right)\right].
\end{equation}
Where n presents the time, N is the number of the coupled map,
$\epsilon$ corresponds to internal and external-coupling,
$\epsilon_{\alpha}^{'}(\epsilon_{\alpha}^{''})$ stand for coupling
parameters and $\Phi$ is a 1D map in this model. For
$\epsilon_{\alpha}^{'}(\epsilon_{\alpha}^{''})\rightarrow 0$,
there is no coupling at all, hence, local neighborhoods have no
influence on the behavior of the coupled map lattices. Using
Bose-Mesner algebra language, we can write:
\begin{equation}
x_i(n+1)=\left[\epsilon
\sum_{j}A_{ij}^{'}\Phi\left(x_j(n)\right)+(1-\epsilon)\Phi(\sum_{j}
A_{ij}^{''}x_j(n))\right],\quad i,j=1,..,N
\end{equation}
 where
\begin{equation}
{A}^{'}=\sum_{\alpha}\frac{\epsilon_{\alpha}^{'}}{k_{\alpha}}A_{\alpha},
\quad
{A}^{''}=\sum_{\alpha}\frac{\epsilon_{\alpha}^{''}}{k_{\alpha}}A_{\alpha}.
\end{equation}
 $A^{'}$ and $A^{''}$ are the elements of Bose-Mesner Algebra.
 $\epsilon_{\alpha}^{'}$ and $\epsilon_{\alpha}^{''}$ with the same
relation in the associated
 schemes are the coupling constant in the coupled map lattice
  \be\sum _{\alpha=0}^{d}\epsilon_{\alpha}^{'}=1, \quad\sum
_{\alpha=0}^{d}\epsilon_{\alpha}^{''}=1\ee
  Dynamical system  symmetry consists of permutations which do not
affect the adjacency matrices and elements of Bose-mesner algebra.
 \begin{description}
  \item[a-]  By considering $\epsilon=0$, we can generate
Internal-coupled
  maps as follows:
  \begin{equation}
x_i(n+1)=\Phi\left(
\sum_{j=0}A_{ij}^{'}x_j(n)\right)=\Phi\left(\sum_{\alpha=0}^{d}\frac{\epsilon_{\alpha}^{''}}{k_{\alpha}}
(\sum_{x\in\Gamma_{\alpha}(x_{i}(n))}x(n))\right)
\end{equation}
  \item[b-] By considering $\epsilon=1$, we can generate
External-coupled
  maps as follows:

\begin{equation}
x_i(n+1)=\sum_{j=0}
A_{ij}^{''}\Phi\left(x_j(n)\right)=\sum_{\alpha=0}^{d}\frac{\epsilon_{\alpha}^{'}}{k_{\alpha}}
\left(\sum_{x\in\Gamma_{\alpha}(x_{i}(n))}\Phi(x(n))\right)
\end{equation}
\end{description}
\subsection{Local and globally coupled maps}
We generalize the models $(3.11)$ to the Coupled map lattices with
local-global couplings on a one dimensional lattice of length N
with periodic boundary conditions\cite{S3-1}. These maps follow
Eq. (3.11) if we modify the coupling constants in the following
way:
\be\frac{\epsilon_{\alpha}^{'}}{k_{\alpha}}=\frac{\tilde{\epsilon}_{\alpha}^{'}}{k_{\alpha}}+\frac{\gamma}{N}\quad\quad\quad\quad\quad\alpha=0,1,...,d\ee
where $\tilde{\epsilon}_\alpha$ and $\gamma$ are coupling
parameters. By substituting $\epsilon_{\alpha}$ in (3.11):

$$
 x_i(n+1)=\epsilon \left(1-\sum
_{\alpha=1}^{d}\tilde{\epsilon}^{'}_{\alpha}-\gamma\right)\Phi(x(n)
)+\epsilon
\sum_{\alpha=1}^{d}\frac{\tilde{\epsilon}_{\alpha}^{'}}{k_{\alpha}}\left(\sum_{x\in\Gamma_{\alpha}(x_{i}(n))}\Phi(x(n)
)\right)$$
$$
+(1-\epsilon)\Phi \left[(1-\sum
_{\alpha=1}^{d}\tilde{\epsilon}^{''}_{\alpha}-\gamma)x(n)\right.
\left.+(1-\epsilon)\sum_{\alpha=1}^{d}\frac{\epsilon_{\alpha}^{''}}{k_{\alpha}}(\sum_{x\in\Gamma_{\alpha}
(x_{i}(n))}x(n))\right]$$
\begin{equation}
+\frac{\gamma}{N}\left[\epsilon
\sum_{\alpha=0}^{d}(\sum_{x\in\Gamma_{\alpha}(x_{i}(n))}\Phi(x(n)
))+(1-\epsilon)\Phi(\sum_{\alpha=0}^{d}(\sum_{x\in\Gamma_{\alpha}(x_{i}(n))}x(n)))\right]
\end{equation}

\subsection{Synchronization of coupled maps}
Synchronization of two (or more) chaotic dynamical systems
(starting with different initial conditions) means that their
chaotic trajectories remain in step with each other during the
temporal evolution. Complete synchronization was popularized after
the seminal papers of Pecora and Carroll \cite{S1-3,S1-4}. \\
Complete synchronization in the coupled map means the existence
of an invariant one-dimensional sub-manifold $x_1=...=x_N$ or
$\left(x_1(n)=...=x_N(n)\; \Leftrightarrow\;
x_1(n+1)=...=x_N(n+1)\right)$.\newpage As it was discussed in
section 3, the matrices ${A}^{'} $ and ${A}^{''} $ appearing in
coupled map (3.11) are semi-simple, and hence possess  the minimal
idempotent $E_0,..., E_d$,  where
\begin{equation}
E_0 =\frac{1}{N}J_{N}=\frac{1}{N} \left(%
\begin{array}{cccc}
  1 & .. & 1 \\
\end{array}%
\right)\left(%
\begin{array}{c}
  1 \\
  . \\
  1 \\
\end{array}%
\right)
\end{equation}
projects coupled dynamical system on synchronized state. The
remaining ones, on the other hand, will project it on the
transverse modes. By choosing $i=j$ and considering $(3.12)$ the
generic model of coupled map lattice Eq. (3.11) is reduced to
$x_i(n+1)=\Phi(x_{i}(n))$. Thus, at synchronized state, the
behavior of coupled map lattice is studied by the behavior of
single dynamical element.
 Here in
this article we study the synchronization of the coupled map
lattices with different coupling topologies such as the global
coupling, the nearest neighbor coupling and the coupling varying
with distance etc.\\
In this paper we are interested in explaining the generic model
based on the hierarchy of one and many-parameter chaotic maps
which are introduced in our previous papers\cite{S3-3,S3-4,S3-5}.
 One-parameter families of
chaotic maps of the interval [0, 1] with an invariant measure can
be defined as the ratio of polynomials of degree N (See
\cite{S3-2} for more detail):
\begin{equation}
\Phi_{N}^{(1,2)}(x,\alpha)=\frac{\alpha^2F}{1+(\alpha^2-1)F},
\end{equation}
If F is substituted  with chebyshev polynomial of type one
$T_{N}(x)$, we will get $\Phi_{N}^{(1)}(x,\alpha)$ and if F is
substituted  with chebyshev polynomial of type two $U_{N}(x)$, we
will get $\Phi_{N}^{(2)}(x,\alpha)$. In hierarchy of elliptic
chaotic maps F is substituted by  Jacobian elliptic functions of
{\bf cn} and {\bf sn} types. Also, we present some examples of
discrete-time one dimensional dynamical system in Appendix {\bf
B}.

\section{ Invariant measure at synchronized state}
\setcounter{equation}{0}  The measure which describes the ergodic
properties with respect to the typical initial conditions is
usually called SRB measure\cite{S3-6}. The difficulty in proving
rigorously that a given coupled map lattice exhibits
spatio-temporal chaos lies in the finding of such a SRB measure,
which has the following important properties:
\begin{description}
    \item[a-] The measure is invariant with respect to   symmetry
transformation of N-dimensional dynamical system, namely it
corresponds to the scalar representation of its symmetry group.
    \item[b-] The measure is smooth along unstable eigendirections in
the phase space.
    \item[c-] The measure has strong ergodic properties, including
mixing and positive KS-entropy\cite{S3-7}.
\end{description}
Symmetric transformation are kind of transformations which leave
the adjacency matrices invariant.
 As we will see later, by
choosing the one-dimensional maps with an invariant measure, such
as logistics map or the ones introduced by authors in
\cite{S3-3,S3-4}, the coupled maps  can display the invariant
measure at synchronized state. The corresponding Forbenious-Perron
(FP) equation for N-coupled map $(3.9)$ is shown:
$$
\mu(x_1(n+1),...,x_N(n+1)) =\int dx_1 ...\int dx_N
\delta\left(x_1(n+1)-F_1(x_1(n),...,x_N(n))\right)...$$
\begin{equation}
 \delta(x_N(n+1)-F_N(x_1(n),...,x_N(n)))\mu(x_1(n),..,x_N(n))
 \end{equation}
Considering the above mentioned properties of the invariant
measure, it is obvious that, at synchronized state it should have
the following form:
 \begin{equation}
\mu(x_1,...,x_N)=\delta (x_2-x_1)....\delta(x_N-x_1)\mu(x_1),
\end{equation}
where $\mu(x_1)$ corresponds to the invariant measure of
one-dimensional map defined in one-dimensional invariant manifold
$x_1=x_2=...=x_N $. By exercising the synchronization condition in
Eq. $(4.1)$, we have:
 $$\mu(x_1(n+1),...,x_N(n+1))=\int dx_1 ...\int dx_N
 \delta(x_1(n+1)-F_1(x_1(n),...,x_N(n)))...
 $$$$\delta(x_N(n+1)-F_N(x_1(n),...,x_N(n)))\delta
(x_2(n)-x_1(n))....\delta(x_N(n)-x_1(n))\mu(x_1)$$
$$=\int dx_1
\delta(x_1(n+1)-F_1(x_1(n),..,x_1(n)))...
 \delta(x_N(n+1)-F_N(x_1(n),..,x_1(n)))\mu(x_1)$$ $$ =
\delta(x_2(n+1)-x_1(n+1))..
 \delta(x_N(n+1)-x_1(n+1))\int dx_1\delta
(x_1(n+1)-F_1(x_1(n),..,x_1(n)))\mu(x_1(n)),$$ Now, if the
one-dimensional map $x(n+1)=F(x_1(n),...,x_1(n))$ possesses the
invariant measure $\mu(x_1(n))$, we can write:
\begin{equation}
\mu(x(n+1))=\int\delta(x(n+1)-F(x_1(n),...,x_1(n))d\mu(x_1),
\end{equation}
Then, we have
\begin{equation}
 \mu(x_1(n+1),...,x_N(n+1))=\delta
(x_1(n+1)-x_2(n+1))....\delta(x_N(n+1)-x_1(n+1))\mu(x_1(n+1)),
\end{equation}
 We have already derived analytically invariant
measure for One-parameter families of chaotic maps $(3.19)$ by
using arbitrary values of the control parameter $\alpha$ and for
each integer values of $N$.
\begin{equation}
\mu_{\Phi_N^{(1,2)}(x,\alpha)}(x,\beta)=\frac{1}{\pi}\frac{\sqrt{\beta}}{\sqrt{x(1-x)}(\beta+(1-\beta)x)},\quad
\beta >0
\end{equation}
 It should be mentioned that reader may refer to our previous
papers for derivation and the relation between the control
parameter and $\beta$ \cite{S3-4,S3-5,S3-6}.
 \section{Stability analysis at synchronized state }
\setcounter{equation}{0}
 In this section we present the stability
analysis of the coupled map lattice synchronized state of fix
point type (i.e, one-dimensional map $(3.19)$ at synchronized
state posses fix point) and synchronized chaotic type (
one-dimensional map at synchronized state is chaotic and posses an
invariant measure). This section also studies Li-Yorke type of
chaos as one-synchronized state becomes unstable.
\subsection{Stability analysis by Lyapunov exponent spectra}
\setcounter{equation}{0} The stability of the coupled map can be
assessed by computing its Lyapunov exponent spectrum. The spectrum
of Lyapunov exponents of coupled map lattice with respect to the
synchronization state can be evaluated in a way similar to that of
one dimensional local maps \cite{S4-1}.
 At synchronized state,
the Lyapunov exponents $\Lambda_{\beta}$ of N-dimensional
dynamical system  described by the map $(3.11)$ can be obtained by
perturbing the  state $x_i(n+1)$ infinitesimally around the
synchronized state $(x_1(0)=x,...,x=x_N(0))$; one can show that
$\delta x(m+1)$, the corresponding perturbed state at the time
$m$, becomes \cite{S3-6}: \be \delta
x_{i}(m+1)=\sum_{j}(\frac{\partial x_{i}(m+1)}{\partial
x_{j}(m)})_{x_{1}=...=x_{N}}\delta x_{i}(m),\ee
 Then, by taking derivative of $x_{i}(m+1)$ with respect to
$x_{j}(m)$, in equation of $(3.11)$,
  and exercising the result in $(5.1)$, we obtain:
\be\delta x(m+1)=\Phi'(x(m))(\epsilon
A^{'}+(1-\epsilon)A^{''})\delta x(m) \ee
 by iterating $(5.2)$ we obtain: \be\delta
x(n)=(\prod_{m=0}^{n-1}\Phi'(x(m)))(\epsilon A^{'}+(1-\epsilon
)A^{''})^{m}\delta x(0), \ee with
\be\lambda_{\beta}=\sum_{\alpha}\frac{\eta_{\alpha}}{k_\alpha}P_{\beta\alpha}\ee
where
$\eta_{\alpha}=\epsilon\epsilon_{\alpha}^{'}+(1-\epsilon)\epsilon_{\alpha}^{''}$.
 By using equation $(3.5)$ and
idempotency property of $E_{\alpha}$, we obtain: $$(\epsilon
A^{'}+(1-\epsilon )A^{''})^m=\sum_{\beta}\lambda_{\beta}^m
E_{\beta},$$ substituting in (5.3):
$$\delta
x(n)=\prod_{m=0}^{n-1}\acute{\Phi}(x(m))\sum_{\beta}\lambda_{\beta}^mE_{\beta}\delta
x(0),$$ where
\\
\be \delta x(n)=\prod_{m=0}^{n-1}\acute{\Phi}(x(m))E_{0}\delta
x(0)+\sum_{\beta=1}^{d}\prod_{m=0}^{n-1}\acute{\Phi}(x(m))\lambda_{\beta}^{m}E_{\beta}\delta
x(0) \ee It should be reminded that in Bose-mesner algebra
perspective, by resorting to projection operator, one could
diagonalize the adjacency matrix, where diagonalization is the
common problem in
the coupled map lattice.\\
$E_{0}\delta x(0)$ is a prolongation which depends on
synchronization state whereas $E_{\beta}\delta x(0) $
$(\beta=1,...,d)$ prolongations depend on transverse modes. The
eigenvalues
$(\prod_{m=0}^{n-1}\acute{\Phi}(x(m))\lambda_{\beta}^m)$ are
called the Floquet (stability) multipliers of the orbit. The
 dynamical system phase space is formed by sum-direct of sub-space
 due to its idempotents project. Thus, for stabilities of
 subspace, the absolute value of Floquet multipliers should be
 less than 1.
  \be
\rho_{0}=\prod_{m=0}^{n-1}\acute{\Phi}(x(m))
\quad\quad\quad\quad\quad\quad\quad\quad
\rho_{\beta}=\prod_{m=0}^{n-1}\acute{\Phi}(x(m))\lambda_{\beta}^{m}\quad\quad(\beta=1,...,d)
\ee
 Note that $\rho_{0}$ is eigenvalue which corresponds to
synchronous periodic orbit while $\rho_{\beta}$ to transverse
states. The eigenvalues of the Floquet (stability) multipliers are
real having only 1 value in magnitude . We will show them in
non-increasing order:
$$
\mid\rho_0\mid \geq \mid\rho_2 \mid\geq ...\geq \mid\rho_d\mid
$$
 As the Eq. $(5.4,6)$ shows, $\rho_{\beta}$ directly
corresponds to $E_{\beta}$ (projection operator). So, by
considering the Eq. $(3.6)$, one can diagonalize adjacency matrix
spontaneously. This simplifies the calculation of $\rho_{\beta}$.
Stability analysis of the coupled map lattice based on the
stability of synchronous periodic orbit  and transverse state,
should be done  in one of these three categories:
\begin{description}
    \item[a-] Both of synchronous periodic orbit and transverse
    states are stable. So, feasible region is studied by Lyapunov exponent and Lyapunov
function.
    \item[b-] Synchronous periodic orbit is unstable, but transverse
states are stable. So, feasible region is studied by Lyapunov
exponent and Lyapunov function.
    \item[c-] Both synchronous periodic orbit and transverse states
are unstable. So, stability analysis is studied by Li-yorke
theorem.
\end{description}
By taking the logarithm of power $\frac{1}{m}$ absolute value of
Floquet multipliers, one can study analytically the stability of
dynamical systems with Lyapunov exponent, whereas Floquet
multipliers should be studied numerically. Notice that, $\delta
x(0)$ belongs to corresponding subspace of $E_{\alpha}$, So, the
Lyapunov exponents of N-dimensional dynamical system
$\Lambda_{\beta}$  are defined as:
 \be\Lambda_{\beta}=\lim_{n\rightarrow
\infty}\frac{1}{n}\ln\left(\frac{\|\delta x(n)\|}{\|\delta
x(0)\|}\right)=\ln\mid \lambda_{\beta}\mid+\lim_{n\rightarrow
\infty} \frac{1}{n}\sum_{m=0}^{n} \ln |\Phi'(x(m))|\ee By
considering one-dimensional ergodic map as a dynamical element,
the last part of the above-mentioned Eq. $(5.7)$ is reduced to:
 \be \lim_{n\rightarrow
\infty} \frac{1}{n}\sum_{m=0}^{n} \ln
|\Phi'(x(m))|=\int_{0}^{\infty}d(\mu(x))\ln
|\Phi'(x(m))|=\lambda_{L}(\Phi)\ee where $\lambda_{L}$ shows
Lyapunov exponent of one dimensional map. Therefore, the Lyapunov
exponents of N-dimensional dynamical system is defined by:
 \be\Lambda_{\beta}=\ln\mid
\lambda_{\beta}\mid+\lambda_{L}(\Phi).\ee If in the Eq. $(5.9)$
"$\beta$" is substituted by "0", the equation would be reduced to:
 \be
\lambda_{0}=\sum_{\alpha}\frac{\eta_{\alpha}}{k_{\alpha}}P_{0\alpha}=1
\ee hence, $\Lambda_0=\lambda_{L}(\Phi)$. According to item
$\bf{b}$, for stability of transverse modes it is necessary to
$\Lambda_{\beta}<0$, ($\beta=1,...,d$) therefore:
 \be
|\sum_{\alpha}\frac{\eta_{\alpha}}{k_{\alpha}}P_{\beta\alpha}|\leq
e^{-\lambda_{L}(\Phi)}\ee as it was discussed
$\sum_{\alpha=0}^{d}\eta_{\alpha}=1$.
 Therefore, synchronized
state makes the coupling constants meet the following inequality,
condition:
\begin{equation}
1-e^{-\lambda_{L}(\Phi)}\leq \sum_{\alpha=1}^d
(k_{\alpha}-P_{\beta \alpha})\frac{\eta_{\alpha}}{k_{\alpha}}\leq
1+e^{-\lambda_{L}(\Phi)}\quad\quad\quad\quad\beta=1,..,d
\end{equation}
By considering that the multiples of $\eta_{\alpha}$ are positive,
the gradient of this plans is in the first region. So, the Eq.
$(5.12)$ defines a hyper-parallelogram in hyperspace formed by
$\eta_{\alpha}$'s. If $d=2$, this region would be a parallelogram
to the first region. Considering the Eq. $(5.12)$ as a semi-linear
one, we should use linear function and its  minimal condition to
distinguish the feasible region.  The region is defined as:
$$k_{\alpha}(1-e^{-\lambda_{L}(\Phi)})\leq \epsilon_{\alpha}\leq
k_{\alpha} (1+e^{-\lambda_{L}(\Phi)}),$$
 in ref.\cite{S3-4} is a subset of
feasible region in Eq. (5.12). According to appendix {\bf A}, we
have $-k_{\alpha}\leq P_{\beta\alpha}\leq k_{\alpha}$ and by
noting that $J=NE_{0}$ and considering Eq.(3.4) we can write: \be
(N-\sum_{\beta=0}^{d}
P_{0\beta})E_{0}=(\sum_{\alpha=1}^{d}\sum_{\beta=0}^{d}P_{\alpha\beta})E_{\alpha}\ee
 because $E_{\alpha}^{^{,s}}$ ($\alpha=0,...,d$) are independent,
 we may conclude:
 $$\sum_{\beta=0}^{d}P_{0\beta}=N,\quad\sum_{\alpha=1}^{d}P_{\alpha\beta}=0$$
 according (3.7), $P_{\alpha0}=1$
 and by using relation (5.12)
 \be
k_{\alpha}(1-e^{-\lambda_{L}(\Phi)})<\sum_{\alpha}(k_{\alpha}-P_{\beta\alpha})\eta_{\alpha}<k_{\alpha}(1+
 e^{-\lambda_{L}(\Phi)})\ee
  which confirms the result of Ref. \cite{S3-4}.
  To investigate the
abolition of asynchronization and also in order to calculate the
scaling exponent of its suppression, we linearize the recursion
map (3.11) near the fixed direction (synchronization state) of
this map: \be lim_{n\longrightarrow\infty}x_{i}(n)=x \ee This
leads us to write $ x_{i}(n)=x+\delta x_{i}(n)$. As expected, the
maximum eigenvalue corresponds to synchronization state, giving
the power scaling exponent of dynamical system\cite{S4-23}. Then,
\be \mid
\lim_{n\longrightarrow\infty}\frac{x_{k}}{x_{l}}-1\mid=e^{-n\overline{\rho}}
\ee by regarding Eq. $(5.16)$ and by noting that $\rho_{SLF}$ is
the second largest Floquite multiplier, the scaling exponent of
suppression of the asynchronization, $\overline{\rho}$ is defined
as: \be
\overline{\rho}=\ln\mid\frac{\rho_{max}}{\rho_{SLF}}\mid=-\ln(
\mbox{second largest eigenvalue of
A})=-\ln\mid\lambda_{SLE}(A)\mid \ee  where $\lambda_{SLE}(A)$
means second largest eigenvalue of $A$.  We can describe the
transition probability with respect to synchronized state by:
$$\parallel x_N(n)-x_{syn}\parallel \simeq
e^{-n\lambda_{syn}}$$ Transition probability changes in each time
step  $(n \rightarrow n+1)$,  and takes positive values if
$\epsilon>0 $. The smaller the second largest eigenvalue modulus,
the faster the Lyapunov exponent converges to its equilibrium
distribution (Synchronization state). This problem can be
formulated as a convex optimization problem, which can in turn be
expressed as a semi-definite program \cite{S4-21}. Restoration of
synchronization time $t_{RS}$ is a time that in it system transits
from synchronization state into asynchronization state. It is in
fact the inverse of $\overline{\rho}$
 \be
t_{RS}=\frac{1}{\bar{\rho}}=-\frac{1}{\ln\mid\lambda_{max}(A)\mid}.\ee
From $(5.18)$ it is clear that, Synchronization time t is
independent of Lyapunov exponent of map \cite{ S4-232}.
\subsection{Stability analysis with Lyapunov function}
 Lyapunov function\cite{S4-3} is the generalization of an energy-like function
that is decreased along trajectories. If the lyapunov function
exists, then closed orbit will be forbidden. The formation of
synchronization can be studied through the Lyapunov function. It
states that for all asymptotically stable systems, there exists a
Lyapunov function whose derivative along the trajectories of the
system takes a negative value. To study the stability region of
the coupled map lattices, one can define lyapunov function in
terms of $E_{\alpha}$ as follows:
 \be
 \sum_{\alpha \neq o}E_{\alpha}=I_{N}-\frac{J_{N}}{N}=E_{AS}
 \ee
So, we can define the Lyapunov function as:
 \be
V(n)=\sum_{i,j}^{N}x_{i}(n)(E_{AS})_{ij}x_{j}(n)=\vec{x}^t(n)(I_n-\frac{J_n}{n})\vec{x}(n)
\ee Lyapunov function takes "0" value at synchronized state and in
step $n$ it becomes:
 \be V(n)=\sum_{i=1}^{N}
x_{i}(n)^{2} -\frac{(\sum_{i=1}^{N} x_{i}(n))^{2}}{N} \ee Clearly
$V(n) \geq 0$ and the equality represents the exact synchronized
state. For the asymptotic global stability of the synchronized
state, Lyapunov function must satisfy the following condition in
the region of stability $ V(n+1)< V(n)$ \cite{S4-5}. we can write
  $\overrightarrow{x}$ state in the following form
 $\vec{x}(n)=\vec{x_s}(n)+\delta \vec{x}(n)$
where
  \be \vec{x_s}(n)= \left(\begin{array}{c}
  1  \\
\vdots \\
 1 \\
\end{array}\right)x(n)
\ee   $\vec{x_{s}}(n)$ is the synchronized state. By considering
$(4.9)$, $ V(n)=\delta \vec{x}^t(n)\delta \vec{x}(n) $,
substituenting $\vec{x}(n)$ (corresponding to $\vec{x}(n+1)$) in
(3.11) and by noting that $\delta\vec{x}(n).\vec{x_{s}}(n)=0$,
\begin{equation}
\delta \vec{x}(n+1)=\acute{\Phi}(x(n))\left[\epsilon
A^{'}+(1-\epsilon)A^{''}\right]\delta \vec{x}(n)
\end{equation}
according to $(5.21)$, we can write the ratio of the Lyapunov
function in step $n+1$ to Lyapunov function in step $n$
\begin{equation}
 \frac {V(n+1)}{V(n)}=\frac{\delta \vec{x}^t(n+1)\delta
\vec{x}(n+1)}{\delta \vec{x}^t(n)\delta \vec{x}(n)} =\frac{\delta
\vec{x}^t(n)\left[\epsilon A^{'}+(1-\epsilon)A^{''}\right]^2\delta
\vec{x}(n)\acute{\Phi}^2(x(n))}{\delta \vec{x}^t(n)\delta
\vec{x}(n)}
\end{equation}
where according to reference \cite{S3-4}, $\acute{\Phi}(x)$
 with N=even equal to $\frac{N^{2}}{a^{2}}$. By noting
$ \mid \lambda_1 \mid \geq ...\geq \mid \lambda_d \mid $,
$A=\epsilon A^{'}+(1-\epsilon)A^{''}$ and according to appendix
{\bf A}, one can write
\begin{equation}
\mid \delta x^{t}(n)A^2\delta x(n)\mid\;\;\leq \;\;\parallel\delta
x(n)\parallel^{2} max (\mid\lambda_{A}\mid^{2})
\end{equation}
 therefore,
\begin{equation}
\frac {V(n+1)}{V(n)}\leq max\{\mid \acute{\Phi}(x(n))\mid^{2}:x(n)
\in [0,1]\}(\mid \lambda\mid^2)_{max}\leq 1
\end{equation}
The Eq. $(5.26)$ is confirmed by the results of studies that have
already been obtained in the literature \cite{S4-5}. The above
mentioned result is converted to Lyapunov exponent method in
section $5.1$. For this purpose, we write the log of geometric
mean of square root of Lyapunov function as: \be
\lim_{n\longrightarrow\infty}\frac{1}{n}\ln\left(\sqrt{\prod_{m=0}^{n-1}\mid\frac{V(m+1)}{V(m)}\mid}\right)=
\lim_{n\longrightarrow\infty}\frac{1}{n}\prod_{m=0}^{n-1}\acute{\Phi}(x(n))^{2}\sqrt{\frac{\delta
\vec{x}^t(0)A^{2n}\delta \vec{x}(0)}{\delta \vec{x}^t(0)\delta
\vec{x}(0)}} \ee by choosing $\delta x(0)$ as the idempotent of
$E_{\alpha}(\alpha=1,...,d)$, obviously, the righthand side of
$(5.33)$ equal to $ \lambda_{L}(\Phi)+\ln\mid\lambda_{\alpha}\mid
$, otherwise, if $\delta \vec{x}^t(0)E_{max}\delta
\vec{x}(0)\neq0$, it becomes $
\lambda_{L}(\Phi)+\ln\mid\lambda_{max}\mid$. Where
$AE_{max}=\lambda _{max}E_{max}$. It means that our results
confirm Lyapunov exponent approach in subsection $5.1$.
\subsubsection{Stability of periodic orbit} In order to
prove the stability of synchronized fix point we can rewrite the
Lyapunov function and Lyapunov spectra at a synchronized state. By
refereing to the definition of the coupled map lattice lyapunov
spectra $(5.7)$ at fix point $x^{\star}$,
 $\Lambda_{\beta}(x^{\star})=\ln\mid
\lambda_{\beta}\Phi^{'}(x^{\star})\mid $. Both $\lambda_{\beta}$
and $\mid\Phi^{'}(x^{\star})\mid$ are smaller than $1$. Therefore,
coupled map is stable. Also by refereing to the definition of the
Lyapunov function  at the synchronized state, we can prove the
stability of fixed point at the synchronized state. We can rewrite
relation (5.22) at fixed point $x_{\star}$ as the following form:
\be \vec{x_{s}}=x^{\star}\left(\begin{array}{c}
  1  \\
\vdots \\
 1 \\
 \end{array}\right) \ee
Where $x^{\star}$ is coordinate of synchronized state. According
to appendix {\bf A}: \be \frac {V(n+1)}{V(n)}\leq \mid
\acute{\Phi}^2(x^{\star}(n))\mid\leq 1 \ee by noting that
$x^{\star}$ is fixed point, then $\acute{\Phi}(x^{\star})$ is
smaller than $1$. So, for the stability of the coupled map the
following condition should be met. \be \frac {V(n+1)}{V(n)}\leq 1
\ee
\subsection{Study of unstable synchronize  and transverse  state with
Li-Yorke method} Here we present the study of the topological
effects on the dynamic behaviors of a coupled complex network in
transition to chaos based on Li-Yorke idea. We considered a
network consisting of nodes that are in non-chaotic states with
parameters in non-chaotic regions before they are coupled
together. It was shown that if these non-chaotic nodes are linked
together through a suitable structural topology , positive
Lyapunov exponents of the coupled network $\Lambda_{\beta}$ can be
generated by choosing a certain uniform coupling strength
$\epsilon^{'}(\epsilon^{''})$, and the threshold for this coupling
strength is determined by the complexity of the network
topology.\\
Li-Yorke theorem \cite{S4-6} clarifies the existence of a snapback
repeller, which implies the existence of chaos. This theory was
generalized by Marotto to higher-dimensional discrete dynamical
systems\cite{S4-7}. By employing Marotto's theorem
 the transition to chaotic procedure is as follows: consider the
N-dimensional difference equation \be\label {liyo}
x(n+1)=\Phi(x(n)); \quad \quad     x(n)\in R^{n}.
 \ee
Suppose that the equation $(5.31)$ has a fixed point $x^*$. This
fixed point $x^*$ is called a snap-back repeller if:
\begin{description}
  \item[a-] $\Phi$ is differentiable in a neighborhood $B(x^{*}, r)$ of
$x^{*}$ with radius $r> 0$,
  such that all eigenvalues of the Jacobian $[D\Phi ]$ are strictly
larger than one in absolute values;
  \item[b-] There exists a point $x(0) \in B(x^{*}, r)$, with $x(0)\neq
x^{*}$, for some integer $m>0$,
 $\Phi^{m}(x(0)) = x^{*}$ and $\Phi^{ m}$ is
differentiable at $x(0)$ with $\det[D\Phi ^{m}(x(0))] \neq 0$.
\end{description}
If Eq. (5.28) has a snap-back repeller, then (\ref{liyo}) is
chaotic in the sense of Li-Yorke. At the synchronized state, the
Lyapunov exponents $\Lambda_{\beta}$ of the N-dimensional
dynamical system are described by $(5.9)$. Now, by considering
$(5.9)$ at fixed point ($x^{\star}$) we have:
 \be\Lambda_{\beta}(x^{\star})=\ln\mid
\lambda_{\beta}\mid+\ln\mid\Phi^{'}(x^{\star})\mid
 \ee
The transition to chaotic state takes place when:
  \be \mid \lambda_{\beta}\mid_{max}\geq
\frac{1}{\mid\Phi^{'}(x^{\star})\mid}
 \ee the second condition is: \be
\det\left[D\Phi
^{m}(x(0))\right]=\left(\prod_{i=0}^{m-1}\Phi'(x(i)))\det(\epsilon
A^{'}+(1-\epsilon
)A^{''}\right)^{m}=\prod_{i=0}^{m-1}\Phi'(x(i))\prod_{\beta=1}^{d}\lambda_{\beta}^{m}\neq0
\ee Therefore, we need to choose coupling constants
$\epsilon_{\alpha}^{'}$ and $\epsilon_{\alpha}^{''}$ so that all
$\lambda_{\beta}$($\beta=0,...,d$) become nonzero. Also, $x(0)$ is
chosen in such a way that satisfies the $\Phi(x(0))=x^{\star}$ and
$\Phi^{'}(x(0))=\frac{N^2}{a^2}\neq0$.\\
As an example for the trigonometric chaotic maps \cite{S3-4}, we
introduce the above mentioned conditions:
\begin{itemize}
  \item For $\Phi^{(1)}_{N}(x,\alpha)\;(N=\mbox{even})$, as it was
discussed, \cite{S3-4} maps have only a fixed point attractor
($x_{\star}=0$)   then
 $x_{0}=1$.
  \item For $\Phi^{(2)}_{N}(x,\alpha)\;(N=\mbox{even})$, it was
  shown, \cite{S3-4} maps have only a fixed point attractor ($x_{\star}=1$)
then $x_{0}=0$.
\item  By choosing odd value of N in $\Phi^{(1,2)}_{N}(x,\alpha)$ , another fix point presented. For which the Li-yorke condition met
too.
\end{itemize}
The ratio of each polynomial volume to the to total defined volume
is equal to "0" \cite{S4-11}. Then
$\prod_{\beta=1}^{d}\lambda_{\beta}^{m}\neq0$ and Li-Yorke chaos
is happened.
\subsection{Kolmogorov-Sinai entropy}
Kolmogorov-Sinai (KS) entropy and Lyapunov characteristic
exponents are two related ways of measuring 'disorder' in a
dynamic system. A definition of them can be found in many
textbooks\cite{S3-6}. The KS-entropy of coupled map lattice can be
assessed by computing Floquet multipliers integerals. It should be
mentioned that computing the KS-entropy for measurable dynamical
system ( such as $(3.11)$ at synchronized state ) is also
possible. The characterization of chaotic time dynamics can be
made through the KS-entropy of the coupled map lattice. The
ergodic choice of one-dimensional map $x(n+1)=\Phi(x(n))$ leads to
the equality of KS-entropy with sum of positive Lyapunov
exponents. Hence, according to Pessin's theorem \cite{S3-2}, the
ergodicity of one-dimensional map $ x(n+1)=\Phi(x(n))$ implies the
ergodicity of symmetric N-dimensional map $(3.9)$ at unstable
synchronized state ( synchronized state is stable for negative
critical exponent $\Lambda_{\beta}$). Obviously, the non-ergodic
choice of $x(n+1)=\Phi(x(n))$ will lead to the non-ergodicity at
synchronized state. The transition from chaotic synchronization
(spatial order with temporal chaos) to non-synchronized states
with positive KS-entropy (spatial and temporal disorder) occurs.
 Finally, comparing the KS-entropy  with the
sum of $\Lambda_{\beta}$ we have:
\begin{equation}
 KS(\Phi_{N-coupled }\quad \mbox {sychronized
})=\sum_{k=1}\Lambda_{\beta}
(\Phi_{N-coupled}\quad\mbox{sychronized}).
\end{equation}
\section{Some example of well known coupled map with corresponding graph}
\setcounter{equation}{0} Globally coupled maps  are one of the
favorite models in the study of spatially extended dynamical
systems \cite{S5-1}. These models of the coupled map lattice
correspond to the complete graph. A complete graph is a graph in
which each pair of graph vertices is connected by an edge. The
complete graph with {\bf N} graph vertices is denoted $K_N$ and
has $\frac{N(N-1)}{2}$ (the triangular numbers) undirected edges
\cite{S5-3}. Globally coupled maps correspond to the trivial
scheme with agency matrices:
\begin{equation}
 A_0=I_N,\quad A_1=I_N-J_N.
 \end{equation}
  By considering $E_{1}=I_{N}-\frac{J_N}{N}$ and
$E_0=\frac{J_N}{N}$ and by using an arbitrary element of Bose
Mesner algebra, the equation $ A=\epsilon A^{'}+(1-\epsilon)A^{''}
$ is reduced to:
\begin{equation}
A=E_{0}+(1-N\frac{\eta_{1}}{k_1}E_{1})
\end{equation}
then eigenvalues of $A$ are $\lambda_{1}=1$ and $\lambda_{2}=1-N
\frac{\eta_{1}}{k_{1}}$. According to (5.18) and (A-2), it is
clear that restoration of synchronization time $t_{RS}$ is reduced
to:
$$t_{RS}=-\frac{1}{\ln\mid1-N
\frac{\eta_{1}}{k_{1}}\mid}$$
 Based on $(5.12)$:   \be
\frac{k_{1}}{N}\left(1-\exp(-\lambda_{L}(x_{n+1}=\Phi(x_n)))\right)\leq
\eta_{1}
\leq\frac{k_{1}}{N}(1+\exp(-\lambda_{L}(x_{n+1}=\Phi(x_n)))) \ee
 Similarly through Lyapunov function
approach the following condition is met too: \be
\frac{k_{1}}{N}(1-\frac{1}{|\max \Phi'(x_n)|})\leq \eta_{1}
\leq\frac{k_{1}}{N}(1+\frac{1}{|\max \Phi'(x_n)|})\ee Coupling
coefficients of the global coupled map (see as an example
\cite{S5-2}) generates the agency matrix following $(6.1,2)$. So,
 their stability analysis, stable region, can be calculated
by relations $(6.3,4)$. Also, in appendix {\bf C} some of the well
known graphs have been presented to simplify the generation of the
new model.
 \section{Conclusion and outlook}
In conclusion, it was found that the variety of the coupled maps
(internal and external) display an identical pattern.
 Moreover, the synchronization capability of various kinds of
coupling schemes were analyzed. We presented the proofs that the
synchronization rate is mainly determined by the second largest
eigenvalue of the Floquet multipliers.\\
In order to determine whether our results in this paper are
generic or they only depend on the specific model, we also studied
several other coupled maps. In future, the study may  cover the
following issues:
\begin{itemize}
    \item By applying the renormalization group theory in coupled dynamical
systems, the scheme picture of model would be presented. As it was
shown, near the synchronized state, different dynamical systems
show the same patterns. This may lead us to this question "do the
different dynamical systems follow the same pattern at
synchronization state? ".
 \item  As it was shown in this study and in authors previous works \cite{S4-231, S4-232}, in nonlinear
 maps, dynamical element represents
 the scheme picture, which may lead us to another interesting question " can we
 attribute the scheme to every nonlinear dynamical systems and coupled
 dynamical system too? "
\end{itemize}
Further, we do hope that our obtained results through this paper
will pave the way for further studies on nonlinear dynamical
systems.
  \renewcommand{\theequation}{I-\arbic{equation}}
\\\\\\  {\Large{\textbf{Appendix A}}}  \\
  \setcounter{equation}{0}
  \renewcommand{\theequation}{A-\arabic{equation}}
It was proved \cite{S6-2}, that if A is a Hermitian matrix, then
we have $ \mid x^{t}Ax \mid \;\leq \;max (\mid \lambda_{A}
\mid)\parallel x\parallel^{2} $ and if x is the largest absolute
eigenvalue then we have, $ max (\mid \lambda_{A} \mid)$, $ \mid
x^{t}A_{\alpha}x \mid\;\leq \;k_{\alpha}\parallel x
\parallel^{2} $,
Thus for adjacency matrix $A_{\alpha}$, we have $max (\mid
\lambda_{A_{\alpha}} \mid)\;\leq \;k _{\alpha} $ and so
$P_{\beta\alpha}\;\leq \;\mid k _{\alpha} \mid$ for
$A=\sum_{\alpha}\eta_{\alpha}A_{\alpha}$:   \be \mid
x^{t}Ax\mid=\sum_{\alpha}\mid\eta_{\alpha}\mid\mid
x^{t}A_{\alpha}x\mid\;\leq\;\sum_{\alpha} k
_{\alpha}\mid\eta_{\alpha}\mid
\parallel x\parallel ^{2}\ee Taking $\sum_{\alpha}
\eta_{\alpha}k _{\alpha}=1$ and by assuming that $\eta_{\alpha}$
is positive, we have: \be\frac{\mid
x^{t}Ax\mid}{\parallel x\parallel ^{2}}\;\;\leq\;\;1\ee \\
{\Large{\textbf{Appendix B}}}\\
We present some examples of the discrete one dimensional dynamical
system with their invariant measure and Lyapunov exponent in order
to simplify the generation coupled map model based on Bose-Mesner
algebra:\begin{enumerate}
    \item Bernuli shift map:\\
    $\phi(x)=\frac{1}{p_k}(x-\sum_{i=1}^{k-1}p_i)$\quad for
\quad$\sum_{i=1}^{k-1}p_{i} \leq x \leq
\sum_{i=1}^{k}p_{i}$\quad \quad\quad (k=2,3,...)\\
 $$\mu=\frac{1}{p_k},
\quad \lambda=-\sum_{i=1}^k p_i\ln{p_i}$$
    \item Generalized Tent map
    $$\phi(x)=\frac{(-1)^{k+1}}{p_k}(x-\sum_{i=1}^{k-1}p_i)+\frac{1}{2}(1+(-1)^{k})\quad
for \quad\sum_{i=1}^{k-1}p_{i} \leq x \leq \sum_{i=1}^{k}p_{i}\;
(k=2,3,...)$$
 $$\mu=\frac{1}{p_{k}},\quad \lambda=-\sum_{i=1}^k p_i\ln{p_i}$$
    \item Gauss map:
$$\phi(x)=\frac{1}{x}-\left[\frac{1}{x}\right],\quad\mu=\frac{1}{(1+x)\ln{2}}, \quad \lambda =\frac{\pi^{2}}{6\log 2}$$
\item Hut map:\\
$$\phi(x)=\frac{1}{2}\left(-1+\sqrt{9-16|x-\frac{1}{2}|}\quad\right),\quad
\mu=x+\frac{1}{2},\quad \lambda =\frac{1}{2}+2 \ln 2-\frac{9}{8}
\ln 3$$
\item Piecewise parabolic map:\\
$$\phi(x)=\frac{1+r-\sqrt{(1-r)^{2}+4r\mid1-2x\mid}}{2r}
\quad  for \quad r\in [-1,1]$$
$$\mu=1+r(1-2x),\quad\lambda=\ln2+\frac{(1-r)^2}{4r}\ln(1-r)-\frac{5r^2-2r+1}{4r}\ln(1+r)+\frac{r}{2}+1$$
\end{enumerate}
 {\Large{\textbf{Appendix C}}}  \\
  \setcounter{equation}{0}
  \renewcommand{\theequation}{C-\arabic{equation}}
In this appendix we present some examples of the graph in order to
simplify the generation of the coupled map model based on
Bose-Mesner algebra.
\begin{description}
\item [a-]\textbf{Strongly regular
graph}\\
A regular graph  $G$ of degree $r$ that is neither empty and nor
complete is called strongly regular if every pair of adjacent
vertices has exactly $u$ common neighbors and every pair of
non-adjacent vertices has exactly $\nu$ common neighbors. The
numbers $r,u$ and $\nu$ are the parameters of the graph.
 It is proved that a regular connected graph G of degree r is
strongly regular if and only if it has exactly three distinct
eigenvalues $\lambda_{1}=r>\lambda_{2}>\lambda_{3}$. If $G$ is a
strongly regular graph with parameters $u$ and $\nu$ then $
u=r+\lambda_{2}\lambda_{3}+\lambda_{2}+\lambda_{3}$, and
$\nu=r+\lambda_{2}\lambda_{3}$ or $$
\lambda_{2,3}=\frac{1}{2}[u-\nu\pm\sqrt{(u-\nu)^{2}-4(\nu-r)}]. $$
then according to (5.18) and (A-2) it is clear that restoration of
synchronization time $t_{RS}$ is:
$$t_{RS}=-\frac{1}{\ln\mid\frac{1}{2}[u-\nu+\sqrt{(u-\nu)^{2}-4(\nu-r)}]\mid}$$
The parameters $N,r,u$ and $\nu$   determine not only each
eigenvalue,but also the
 multiplicities of each eigenvalue. Since $G$ is connected, the
multiplicity
 of $\lambda_{1}$ is $1$. The multiplicities  $m_{2}$ and $m_{3}$ of
$\lambda_{2}$ and $\lambda_{3}$ must sum to $N-1$ and by noting
that the sum of the eigenvalues must be zero, the sum of the
squares of the eigenvalues must be equal to $N\lambda_{1}=Nr$
.Thus $m_{2}$ and $m_{3}$ are
 $$
\frac{1}{2}(\nu-1\mp\frac{(\lambda_{2}+\lambda_{3})(\nu-1)+2r}{\lambda_{2}-\lambda_{3}}),$$
 An example for strongly regular graph is Hamming graphs with
diameter 2. $H(2, q)$ is an SRG $(q^2, 2(q - 1), (q - 2), 2)$ with\\
$$ Spec(H(2, q)) =\left(\begin{array}{ccc}
  \lambda_{1}&\lambda_{2}&\lambda_{3}  \\1&m_{1}&m_{2}
\end{array}\right)=\left(\begin{array}{ccc}
  2(q - 1)& q - 2& -2\\1&2(q - 1)&(q - 1)^2
\end{array}\right)
$$ if $q=2$ then:
$$\frac{1-e^{-\lambda_{L}(\Phi)}}{4}\leq\eta_{1}\leq\frac{1+e^{-\lambda_{L}(\Phi)}}{4},$$$$
\frac{1-e^{-\lambda_{L}(\Phi)}}{2}\leq \eta_{1}+\eta_{2} \leq
\frac{1+e^{-\lambda_{L}(\Phi)}}{2}.
$$
 and if $q>2$ using Eq. $(5.14)$ we find a feasible region
in the following form:
$$\frac{1-e^{-\lambda_{L}(\Phi)}}{q}\leq\eta_{1}+(q-1)\eta_{2}\leq\frac{1+e^{-\lambda_{L}(\Phi)}}{q},
$$$$
\frac{1+(3-2q)e^{-\lambda_{L}(\Phi)}}{q^2}\leq\eta_{1}\leq\frac{1+(-3+2q)e^{-\lambda_{L}(\Phi)}}{q^2}.$$
then according to (5.18) and (I-4) it is clear that restoration of
synchronization time $t_{RS}$: $$t_{RS}=-\frac{1}{\ln(q-2)}$$.
\item [b-]\textbf{Cycle graph}\\
Another family of the N-coupled maps  generated with  cycle
 graph \cite{S5-3}, is introduced as following:
$$
S=\left(\begin{array}{cccc}
  0& 1 & 0 &... \\
  \vdots& 0 & 1 & \ldots \\
& \ldots & 0 & 0
\end{array}\right)
$$
where $S$ is the shift matrix. In cycle graph, adjancy matrices
are introduced in the following form:$$ \hspace{-3cm}A_0=I_N,\quad
A_i=S^{i}+S^{-i},\quad i=1,...,[\frac{N}{2}],\quad
N=\mbox{odd}$$$$ A_0=I_N,\quad A_i=S^{i}+S^{-i},\quad
i=1,...,\frac{N}{2}-1,\quad A_{\frac{N}{2}}=S^{\frac{N}{2}}\quad
N=\mbox{even}
$$
Nearest neighbor coupled map lattice correspond to cycle graph.
 Nearest neighbor coupled map\cite{S5-1} is used as a model in the
simulation of natural phenomena such as clouds, smoke, fire and
water. It is one of the most important research areas in computer
graphics.  By using (C-6,7) and  $A=\epsilon
A^{'}+(1-\epsilon)A^{''}$: $$
A=(1-\eta_{1})I+\frac{\eta_{1}}{2}(S+S^{-1}),
$$
 where
$\eta_{1}=\epsilon\epsilon^{'}_{1}+(1-\epsilon)\epsilon^{''}_{1}$.
Applying Fourier method we can write idempotents:
 $$
 E_{j}=\left(\begin{array}{c}
  1  \\
\omega^{j} \\
 \vdots\\
\omega^{(n-1)j}
\end{array}\right)(1,\omega^{-j},...,\omega^{-(N-1)j})
 $$
 where $\omega=\exp(\frac{2\pi i}{N})$, so eigenvalues are:
 $$
\tau_{j}=1-\eta_{1}+\eta_{1}\cos \frac{2\pi j}{N},
 $$
The stability condition for cycle graph: $$ \eta_{1}\leq
\frac{1+e^{-\lambda_{L}(X=\Phi(x))}}{2(1+\cos\frac{\pi}{N})},\; \;
\mbox{for odd N},\;\;
\eta_{1}\geq\frac{1-e^{-\lambda_{L}(X=\Phi(x))}}{2(1-\cos\frac{2\pi}{N})},
\; \mbox{for every N}
 $$
According to (5.18) and (A-2) it is clear that restoration of
synchronization time:
$$t_{RS}=-\frac{1}{\ln\mid\cos\frac{2\pi}{N}\mid}$$
\end{description}

\end{document}